# Local Structure of Thermoelectric $Ca_3Co_4O_9$

T. A. Tyson and Z. Chen
*Department of Physics, New Jersey Institute of Technology, Newark, NJ 07102*

Q. Jie and Q. Li.
*Condensed Matter Physics and Materials Science Department, Brookhaven National Laboratory, Upton, NY 11973*

J. J. Tu
*Department of Physics, The City College of New York, New York, NY 10031*

## Abstract

We have combined temperature dependent local structural measurements with first principles density functional calculations to develop a three dimensional local structure model of the misfit system $[Ca_2CoO_3][CoO_2]_{1.61}$ (referred to as $Ca_3Co_4O_9$) which has a rock salt structure stacked incommensurately on a hexagonal $CoO_2$ lattice. The local structural measurements reveal a low coordination of Co(2)-O bonds in the rock salt layer with large static structural disorder. The temperature dependence of the Co(1)-Co(1) bond correlations in the $CoO_2$ layer are found to be normal above ~75K and with a very small static disorder component. An anomalous enhancement in the Co(1)-Co(1) correlations occurs at the onset of long-range magnetic order. Density functional computations suggest that the reduction of the coordination of Co(2) is due to the formation of chains of $Co(2)O_x$ in the a-b plane linked to the Ca-O layers by c-axis Co(2)-O bonds. The reduced dimensionality introduced by the chain-like structure in the rock salt layer and high atomic order in the $CoO_2$ layer may enable low thermal conductivity and high electrical conductivity in the respective layers.

# I. Introduction

Utilizing waste heat from ubiquitous power plants and engines in automobiles would provide a way to make more efficient use of the energy in today society. Approximately 70% of the total primary energy is lost in the form of heat which is widely dispersed and thermoelectrics offer the promise to recover much of this lost energy [1,2,3,4]. In addition, solving the problem of cooling the localized heating in computer CPUs can result in significant speed gain as clock frequencies could be driven higher. The efficiency of a thermoelectric solid is found to depend on the material properties via the dimensionless figure of merit, Z T where $Z = \sigma S^2/\kappa$. In this expression $\sigma$ is the electrical conductivity, S (also called the thermopower) is the average entropy per charge carrier divided by the electron charge and $\kappa$ can be written as $\kappa = \kappa_{electron} + \kappa_{lattice}$ the sum of the electronic and lattice contributions to the thermal conductivity. A value of ZT ~ 3 would make solid state cooling economically competitive with compressor type refrigerators. The best materials available to date in practical applications have ZT values near 1.

Controlling the lattice contribution to thermal conductivity can have a significant impact for enhancing ZT. This has been demonstrated in the materials with (1) a large number of atoms in the unit cell, (2) heavy atoms, (3) systems with large average coordination per atom (4) systems with cage-like structures (such as the skutterudites and clathrates) in which weakly bound atoms possess large thermal motion that results in reduction of lattice thermal conductivity. Understanding the structure of thermoelectrics is crucial to determine how these materials can be optimized for device applications.

The ideal thermoelectric will conduct electrical current like a metal yet have a low thermal conductivity. This concept of a phonon glass/electron crystal (PGEC) was introduced [5] for bulk materials, on the other hand, the idea of a low dimensional system was also considered [6]. As the dimensions of a system decreases new variable becomes available ("system size") to tune ZT independent of the Wiedemann-Franz constraints connecting the electronic thermal conductivity and electron transport. Also as the dimensions are reduced from 3D to 2D to 0D (quantum dots), new phenomena (quantum confinement of electrons) emerge. In addition, the introduction of interfaces in layered materials enhance electron and phonon scattering.

Transition metal oxides are being studied for high temperature application because of their stability. The system $[Ca_2CoO_3][CoO_2]_{1.61,}$ with a ZT value which is estimated to be over 1 at high temperature is currently being extensively investigated and belongs to a class of thermoelectrics with Co in a triangular lattice [4]. This material has a complex structure, called a misfit structure (see Refs [7, 8, 9, 10]), in which a square $CaCoO_3$ layer of NaCl (rock salt) type structure is stacked on hexagonal lattice of $CoO_2$ (with Co in a triangular configuration). A central structural feature of these materials is the split-site disorder at the rock salt Co and O positions and modulation of the Co and O position in both layers [9,10]. X-ray diffraction (XRD) measurements reveal that the split site disorder O position in the rock salt layer exhibits oxygen vacancies. This is supported by iodometric titration experiments which indicate a average Co lower valence than that expected for a fully oxygenated sample when synthesized in air or a reducing environment [11]. As will be seen below, local structural measurements will provide insight into the disorder about Co in the two distinct sites in this system.

Stacking of a triangular lattice with a square lattice results in significant distortions at the interface and the b-axis dimension in each layer which are not commensurate. The structure is abbreviated as $Ca_3Co_4O_9$. This system is a strongly correlated oxide with many intriguing properties (see Refs. [12]). The system behaves like an insulator below ~63K and a Fermi liquid between ~63 K and ~143 K. The transport property of this materials is like a weak metal (incoherent regime) up to ~510 K. It is believed that a high spin to low spin transition on Co occurs at ~ 400 K. Strong pressure and strain dependence is seen in the electrical conductivity. µSR measurements suggest the onset of an incommensurate spin density wave below 100 K with short range order followed by full long range order at 27 K [13].

Although electronic structure calculations [14], soft x-ray spectroscopic studies [15] and Co K-Edge near absorption studies [6] have been carried out on $Ca_3Co_4O_9$ no local structural measurements on this system has been conducted, to our knowledge, to ascertain the role of short-range structural features on the thermoelectric properties is this system.

In this work, we have measured the temperature dependent local structure about the Co sites in $Ca_3Co_4O_9$ by Co K-Edge x-ray absorption spectroscopy (XAFS). Density functional (DFT) calculations were used to qualitatively determine the optimized 3D structure in a 1-cell approximation.

## II. Experimental and Computational Methods

A polycrystalline sample of $Ca_3Co_4O_9$ was prepared by solid state reaction. X-ray diffraction measurements reveal that is was single-phase. In addition, resistivity measurements and near edge x-ray absorption measurements all show that the sample was consistent in oxygen content with those prepared in previously work [8]. An x-ray absorption sample was prepared by grinding and sieving the material (500 mesh) and brushing it onto Kapton tape. Layers of tape were stacked to produce a uniform sample for transmission measurements with a jump (difference between the background and post-edge region of the near threshold spectrum) $\mu t \sim 1$. Spectra were measured at the NSLS beamline X23B at Brookhaven National Laboratory. Measurements were made on warming from 17 K. Two to four scans were taken at each temperature. The uncertainty in temperature is $\sim 0.1$ K. A Co foil reference was utilized for energy calibration. The reduction of the x-ray absorption fine-structure (XAFS) data was performed using standard procedures [16]. The energy $E_0$ (k=0) was chosen to be 7726 eV. Data analysis was carried out as in Ref. [17]. Representative absorption fine-structure data at 270 K are shown in Fig 1(a). Data over the k-range $2.5 < k < 14.8$ Å$^{-1}$ was used and fits in R-space (described below) were carried out over the range $0.8 < R < 3.0$ Å with $S_0^2$=0.86 (accounting for electron loss to multiple excitation channels). A typical R-Space fit is shown in Fig. 1(b). In this figure note that positions of the peaks are shifted to lower R values due to the atomic phase shifts. The accurate atomic position and thermal factors are obtained by fitting to data to the complex phase and amplitude functions of an approximate model of the local structure.

The first two shells about Co are studied in detail (the average Co-O shell and the Co(1)-Co(1) shell in the $CoO_2$ layer). To treat the atomic distribution functions on equal footing at all temperatures the spectra were modeled in R-space by optimizing the integral of the product of the radial distribution functions and theoretical spectra with respect to the measured spectra. Specifically, the experimental spectrum is modeled by, $\chi(k) = \int \chi_{th}(k,r) 4\pi r^2 g(r) dr$ where $\chi_{th}$ is the theoretical spectrum and g(r) is the real space radial distribution function based on a sum of Gaussian functions ($\chi(k)$ is measured spectrum) [18] at each temperature. Theoretical spectra for atomic shells [19] were derived from a cluster model of the average structure crystal structure. The coordination numbers of the average Co-O shell ($<N_{Co-O}>$) was determined by fitting while that of the Co(1)-Co(1) shell held fixed. The temperature dependent widths ($\sigma^2 = \langle (R - \langle R \rangle)^2 \rangle$ giving the mean squared relative displacement of a bond) were determined for both shells. For the Co(1)-Co(1) shell, the extracted widths were fit (above 75K) to a simple Einstein model $\sigma^2(T) = \sigma_0^2 + \frac{\hbar^2}{2\mu k_B \theta_E}\coth(\frac{\theta_E}{2T})$ [20], where $\mu$ is the reduced mass for the bond pair and a parameter $\sigma_0^2$ represents the static disorder. This simple model represents the bond vibrations as harmonic oscillations of a single effective frequency proportional to $\theta_E$. It provides an approach to characterize the relative stiffness of the bonds. In summary, for the Co(2,1)-O shell the distance, coordination number (N), width ($\sigma$) and position parameters were extracted while the coordination number was held fixed for the longer distance Co(1)-Co(1) shell of the $CoO_2$ layer.

Local spin density functional calculations in the projector augment wave approach [21] were carried out using a 42 atom cell as done in Ref. [8]. Here, however, full optimization in P1 symmetry (no imposed symmetry) was conducted for both lattice parameters and atomic positions and the LDA+U approximation was implemented to obtain the fully relaxed structure. The structure was optimized so that forces on each atom were below 5 x 10$^{-4}$ eV/Å. Forces along x, y and z direction were computed for all 42 atoms for 0.03 Å displacements. The local atomic structure was averaged to produce distinct groups (Co(1), Co(2), Ca(1), Ca(2) and distinct oxygen positions of the misfit structure). The computed force constants are unchanged with respect to reductions in the displacements used in the so-called frozen phonon calculations.

## III. Results and Discussion

In Fig. 1(b) we show a representative fit of the Fourier transforms of the XAFS data (structure function) taken at 270 K. The structure function has two peaks corresponding to the average Co(1,2)-O distribution and the Co(1)-Co(1) distribution. The latter peak can be identified by the unique Co(1)-Co(1) distance from XRD measurements. With these two peaks, the structural order in the rock salt and $CoO_2$ layer can be independently assessed by comparing the temperature dependence of the derived structural parameters ($\sigma^2$) for both the Co(1,2)-O and Co(1)-Co(1) bond distributions.

In Fig. 2, we show the coordination number N, and distribution width for the Co(1,2)-O shell (right y-axis). This peak contains both the Co(1)-O and Co(2)-O bonds. Although the theoretical coordination of both sites is six, the measured coordination is ~ 4.5±0.1. The width of the distribution can be written at $\sigma^2$ (thermal) + $\sigma^2$ (static) as mentioned above  The simplest interpretation of the weak temperature dependence is that the static distortions dominate over the entire temperature range studied. Low coordination in XAFS analysis also occurs when distance between the bond pair is large as the scattering amplitude drops off as $1/R^2$. In addition, coupled motion is reduced for atoms which are physically further apart. (Note that a contribution to the reduction may also come from the oxygen deficiency at the O sites for samples produced in air as mentioned above [13]. This is estimated to be 14% vacancies on the O site in the rock salt layer giving an upper limit of 0.3 in reduction of the coordination.) The absence of the long bonds in the Co-O contribution is manifested as a low Co-O average distance of 1.90 Å (Table I). The low coordination number and weak temperature dependence thus point to the same high static disorder in the rock salt layer with a broad range of distances. Since this shell contains both the Co(1)-O and Co(2)-O pairs, examination of the next shell (Co(1)-Co(1), $CoO_2$ layer ) will enable the us to isolate the locally disordered component.

In Fig. 3(a), we show that temperature dependence of the Co(1)-Co(1) shell in the Co(1)$O_2$ layer, having fixed the coordination number for this shell. The high temperature data above ~ 75 K are well represented by an Einstein model ($\theta_E$ = 450(6) K). (The model was extrapolated to temperatures below 75K for comparison.) The static distortions of the Co(1)$O_2$ system are quite small and imply that this component of the system is not connected with low thermal conductivity. Specifically, the static parameter $\sigma_0^2$ was found to be 0.00080(5) Å$^2$ which is very small compared to the values of the total $\sigma^2$ at the lowest temperatures for this correlation. If the oxygen coordination about the Co(1) sites were highly disordered structurally a small $\sigma_0^2$ parameter for the Co(1)-Co(1) correlation would not be expected. This enables us to associate the disorder mentioned above (Co-O shells) with the rock salt layer.

Note also, the large excursion away from this model evident below ~75 K. Raman scattering measurements show phonon hardening at low temperature consistent with this result [22]. Recent µSR experiments reveal a incommensurate spin density wave with short range order onsets at 100 K followed by long range order which stabilizes at 27 K [13]. Further experiments revealed the volume fraction of the magnetic phase in the low temperature spin density wave phase to be 84%. The stability of the spin density wave states [23] at low temperatures in a broader range of three and four layer hexagonal cobaltates and $Na_xCoO_2$ has been well established [13(b)].

In recent work covering the temperature range 2 to 60 K, the field dependent resistance and field dependent magnetization data were combined in a variable range spin dependent hopping model (for a-b plane transport and magnetization) [24]. Writing the resistivity as $\rho(T,H) = \rho_{sp}(T,H) + \rho_{QP}(T)$ for the spin depend (sp) and spin independent quasi particle (QP) contribution to the resistivity, the field dependent magnetization and resistivity data can be used to determine both components of the resistivity. The spin dependent part is found to have an upturn at low temperature while the field independent part of the resistivity exhibits metal like behavior down to low temperatures. Hence the upturn in resistivity, in the system should be thought of as a result of a spin scattering mechanism which dominates over the simple metal like behavior. This is further supported by measurements of the magnetization at low fields in single crystals.

In Fig. 3(b) we compare the temperature dependent $\sigma^2$ parameters (Co(1)-Co(1)) with in-plane (a-c) magnetization data measured on single crystals at low magnetic fields (from Ref. [25]). The magnetization measurements (at 200 Oe) on single crystal revealed the stabilization of the long range incommensurate spin density wave and a ferrimagnetic transition at 19 K as an upturn in the magnetization at ~30K and change in slope of the magnetization, respectively. Hysteresis was observed for M vs. H loops for temperatures below 30K. Note that the dip in the $\sigma^2$ parameters (enhancement of Co(1)-Co(1) correlation) occurs at the onset of the increase in magnetization as temperature is reduced. Thus the Co(1)-Co(1) EXAFS results (local spin-lattice coupling) support the idea that it is the Co(1)$O_2$ layer in which the magnetic ordering occurs.

Combining all of the data for this and previous work we can suggest a model for the transport properties. Electron transport occurs in the Co(1)$O_2$ layers. The increase in resistivity below ~100K is associated with the onset of the short range order incommensurate spin density wave in the Co(1)$O_2$ layer which become a long range ordered incommensurate spin density wave at low temperature. The onset of the long-range ordered magnetic order coincides with enhanced nearest neighbor Co(1)-Co(1) correlations. A residual finite magnetization also occurs with the onset of the long range order.

DFT calculations provide a refined qualitative picture of the 3D local structure. The ground state structure was found and its stability was confirmed by converging the structure with respect to random atomic displacement away from the minimum energy state. The phonons at gamma were determined by a frozen phonon lattice dynamics calculation [26,27]. Three unstable modes found were evaluated by displacing the atoms along the mass corrected eigenvectors of the dynamical matrix to minimize the total energy. The small structural changes found were incorporated into the structural data presented here.

In Table I we compare the structural data obtained from synchrotron XRD measurements [10], the DFT model and the XAFS measurements. In a modulated structure the atomic coordinates of a given species are characterized by [28] $x_i(x_4) = \bar{x}_i + \sum_{n=1}^{\infty} A_i^n \sin(2\pi n x_4) + B_i^n \cos(2\pi n x_4)$ where $\bar{x}_i$ is i$^{th}$ component of the average position of the atom and $A_i^n$ and $B_i^n$ are the Fourier amplitudes which define the modulation functions of the atom. Here $x_4$ is defined as $x_4 = t + \mathbf{q} \cdot \mathbf{x}$. with $\mathbf{q}$ being the modulation vector. The distance between pairs of atoms varies from unit cell to unit cell but all information is contained over the range of t (0 < t < 1). This produces a four-dimensional space. For the diffraction derived structure, atoms with nonzero modulation amplitudes (Table 5 in Ref. [10]), including Co(1) and Co(2), will have a large range of distances (defined by the full range of t parameters) but the average distances are meaningful and can be related to the XAFS results. These are shown in Table I.

Note the similarity in the tabulated structural parameters for the DFT, XAFS and XRD results for the Co(1)-O bonds. Some differences between the methods can be seen by looking at the Co(2)-O shell. Focusing in the XRD results we find three shells (at 1.897, 2.378 and 2.740Å, Table I). we find that excluding the last two Co(2)-O shells yields a coordination number of 4.5 while excluding only the last shell yields a coordination of 5.5. This suggests that weak Co-O correlations between the Co(2)-O atoms in the higher shells may produce the low XAFS coordination. This also explains the short average XAFS bond length. Density functional methods can be used to obtain a more comprehensive qualitative view of the structure.

The density functional calculations of the approximate structure reveal that in the $Ca_2CoO_3$ layer (rock salt layer), chains of $CoO_x$ exist along the a-axis as shown in Fig. 4. This chainlike structure will disrupts the lattice in $Ca_2CoO_3$ (rock salt) layer and possibly reduce the thermal conductivity. Within the approximation, the chain-like structure results in 4-fold coordinated Co(2). High structural order is maintained about the Co(1) sites on the other hand. A seen above (XAFS), the degree of disorder about the Co(1) site is very small and suggest a high degree of symmetry in the coordination of this suite by oxygen. This is not compatible with the existence of oxygen defects in the $CoO_2$ layers although they play a role in the rock salt layer [11].

With the 6 fold coordinated Co(1) sites considered, the average coordination of oxygen about Co is predicted to be ~5.2 in the DFT model (excluding the higher shell). The long bonds are in the a-b plane and span the channels separating distinct chains running along the a-axis. While this coordination number is higher than the observed XAFS results, the observed splitting of the Shell on the Co(2)-O shell into long and short bonds is preserved in a qualitative manner (analogous to the XRD result with three shells). The reduced coordination observed in the measurements (Fig. 2(a)) is possibly the result of disruptions in the chain lengths (with long distances not seen in the XAFS spectrum on average (rock salt layer)). This chain-like structure will yield a large static structural distortions in the Co(1,2)-O distribution.

Analysis of the x, y and z components of the force constants (self-force constants [29]) on all atoms reveals additional anomalies which can add to our qualitative description of the structure. The self force constants, which indicate the force on the isolated atom with respect unit displacements, are all positive indicating that the optimized structure is stable with respect to the displacement of individual atoms. The individual atoms sit in single position wells (no atoms sit on saddle points and there are no "rattling" atoms). The total energy can only be lowered by the cooperative motion of different atoms.

While the Co(1) ions reside in highly symmetric potentials wells with approximately equal force constants for x, y and z displacements (~17 eV/Å$^2$), the Co(2) ions have force constants of ~10 eV/Å$^2$ in the x-y plane and ~40 eV/Å$^2$ along the z-axis. The Co(2) bonded oxygen atoms O2_1 and O2_3 (Fig. 4) exhibit weakened force constants for motion along the x directions with force constants along x, y and z of ~5, ~9 and ~20 eV/Å$^2$. . This can be compared to the Co(1) bound oxygen atoms which have near isotropic values of ~11 eV/Å$^2$. This weakening of the Co(2)-Co bonds will result in suppressed coordination from the Co-O bonds in the rock salt layer.

Summarizing the DFT results we note that, the data at the Co(2) for the model differs from that of the XRD measurement (Table I) in that the there are 4 short and 2 long bonds for DFT model while the XRD data reveal three shells [13]. The qualitative trend is the existence of chain like structure which we have found to be stable and robust with respect to random perturbation of all atomic position in the DFT calculations. The chains results in split long and short Co-bonds about the Co(2) site.

The measurements and modeling provide an atomic scale picture of this system possibly shedding light on thermoelectricity in these materials. The $Ca_2CoO_3$ layer is disordered structurally with characteristic broken chains along the a-axis resulting in reduced dimensionality. In addition, the O2_1 and O2_3 atoms exhibit weak coupling along the x direction. One possibility is that the chain-like features and anomalous O2_1 and O2_3 potential wells will suppress the thermal conductivity in the $CaCoO_3$ layers. By contrast, the higher structural order in the $CoO_2$ layers may enable metal-like electron conductivity. We also, suggest that the chainlike structures in the rock salt layer my possibly serves as a charge reservoir (as in some high temperature superconductors [30] with broken $Cu-O_x$ chains) while transport and magnetic ordering occurs in the $CoO_2$ layer. Calculations on larger unit cells are being carried out to refine this model.

## IV.  Summary

We have combined temperature dependent local structural measurements with first principles density functional calculations to develop a three dimensional local structure model of the misfit system $[Ca_2CoO_3][CoO_2]_{1.61}$ (referred to as $Ca_3Co_4O_9$) which has a rock salt structure stacked incommensurately on a hexagonal $CoO_2$ lattice. The local structural measurements reveal a low coordination of Co(2) in the rock salt layer. The temperature dependence of the Co(1)-Co(1) bond correlations in the $CoO_2$ layer are found to be normal above ~75K and have a very small static disorder component. An anomalous enhancement in the Co(1)-Co(1) correlations occurs at the onset of long-range magnetic order. Density functional computations suggest that the reduction of the coordination of Co(2) is due to the formation of chains of $Co(2)O_x$ in the a-b plane linked to the Ca-O layers by c-axis Co(2)-O bonds. The reduced dimensionality introduced by the chain-like structure in the rock-salt layer and high atomic order in the $CoO_2$ layer enable low thermal conductivity and high electrical conductivity in the respective layers yielding a high thermoelectric figure of merit.

The results suggest that nanoscale structuring of materials through the development of tailored films growth methods will lead to new and novel thermoelectrics with properties which can be tuned to specific applications. Detailed theoretical studies of the effect doping in the rock salt layer (replacement of Ca by Sr or Ba for example) may reveal systems with reduced thermal conductivity in the rock salt layers while still preserving the high atomic order and conductivity in the $CoO_2$ layer.

## Acknowledgements

This work is supported by U. S. Department of Energy Office of Basic Energy Sciences (DOE-BES) Grant DE-FG02-07ER46402 (TAT and ZC) and by DOE-BES grant No. DE-AC-02-98CH10886 (JQ and QL). We thank Prof. C. D. Ling of the University of Sidney for providing us with detailed synchrotron diffraction derived structural parameters for comparison with our x-ray absorption measurements and for providing insightful details on the modulated structure. This research used resources of the National Energy Research Scientific Computing Center, which is supported by the Office of Science of the U.S. DOE under Contract No. DE-AC02-05CH11231.

## Table I. Structural Parameters (Å)*

|            | XRD                              | XAFS (270K)      | DFT**                          |
|------------|----------------------------------|------------------|--------------------------------|
| Co(1)-O    | 1.909 (6)                        | 1.900±0.002      | 1.958±0.070 (6)                |
| Co(2)-O    | 1.879 (2)<br>2.378 (2.67)<br>2.740 (1.33) |                  | 1.848±0.082 (4)<br>2.705±0.155 (2) |
| Co(1)-Co(1)| 2.809 (6)                        | 2.814±0.001 (6)  | 2.844±0.076 (6)                |

\* The numbers presented are averages over the shells. The quantity in parenthesis is the number of atoms in the corresponding shell. The XAFS first shell results represents an average over both the Co(1)-O and Co(2)-O shells. XRD data are from Ref. [10].

\*\*"Errors" in DFT parameters are from standard deviation of bond distances in shell.

# Figure Captions

**Fig. 1.** (a) Extracted Co K-edge XAFS data and (b) Fourier of the transform of the data taken at 270 K. The thin lines are fits to the data (thick lines). Peaks are shifted to lower R-values from real distance by atomic phase shifts.

**(Color online) Fig. 2.** Coordination number and distribution width of Co-O shell. The average first shell oxygen coordination number (right y-axis and circular data points) of Co $<N_{Co-O}>$ is significantly lower than 6. The width (left y-axis), $\sigma$, of the Co(1,2)-O bond distribution as a function of temperate reveal nonstandard temperature dependence.

**(Color online) Fig. 3.** Temperature dependent Co(1)-Co(1) distribution compared with resistivity (a). For the width, $\sigma$, of the in-plane Co(1)-Co(1) distribution ($CoO_2$ layer), the high temperature data above ~ 75 K are modeled well by an Einstein model ($\theta_E$ = 450(7) K) with a static component $\sigma^2$ = 0.00080(5) $Å^2$. Large excursion away from this model are evident below 75 K. Resistivity data [8] show a sharp up-turn where $\sigma^2$ increases showing loss of Co-Co correlation coincides with enhanced resistivity. (b) Comparison of low field (200 Oe) in ab-plane magnetization data for single crystal taken from Ref. [25] with temperature dependent $\sigma^2$ parameters. Note the coincidence of the upturn in the magnetization with the minimum in the dip in the $\sigma^2$ parameters.

**(Color online) Fig. 4.** 3D picture of structure obtained from optimized cell (DFT). Note the $Co(2)O_x$ chains which form parallel to the a-axis in the a-b plane. The dotted blue lines define the crystallographic cells. The blue, red and light blue spheres denote Co, O and Ca atoms, respectively.

**Fig. 1.** Tyson *et al.*

(a)

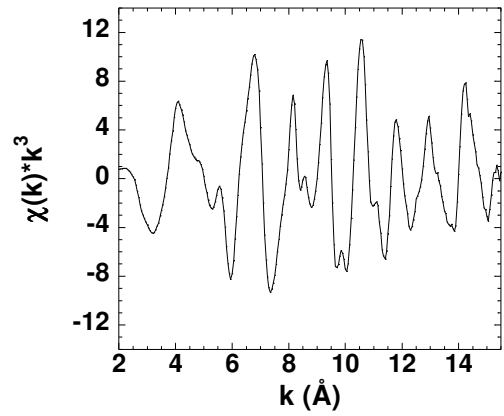

(b)

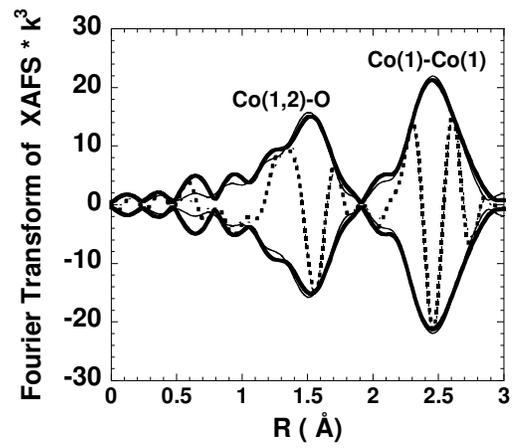

**Fig. 2.** Tyson *et al.*

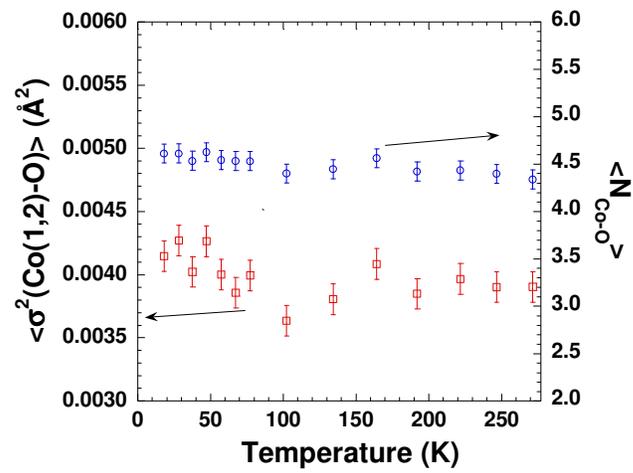

**Fig. 3.** Tyson *et al.*

**(a)**

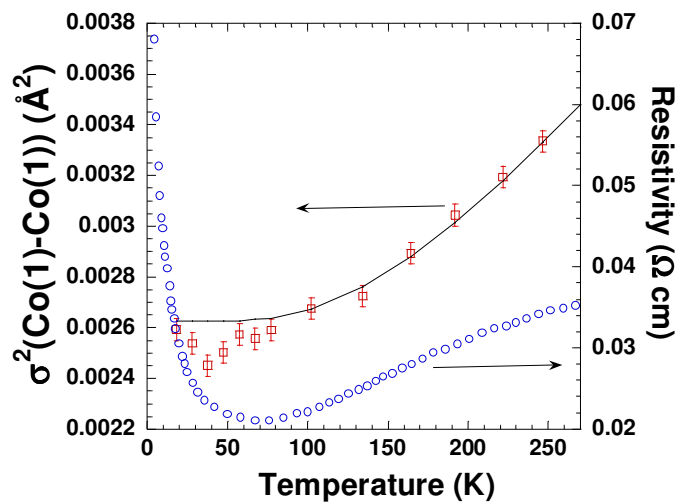

**(b)**

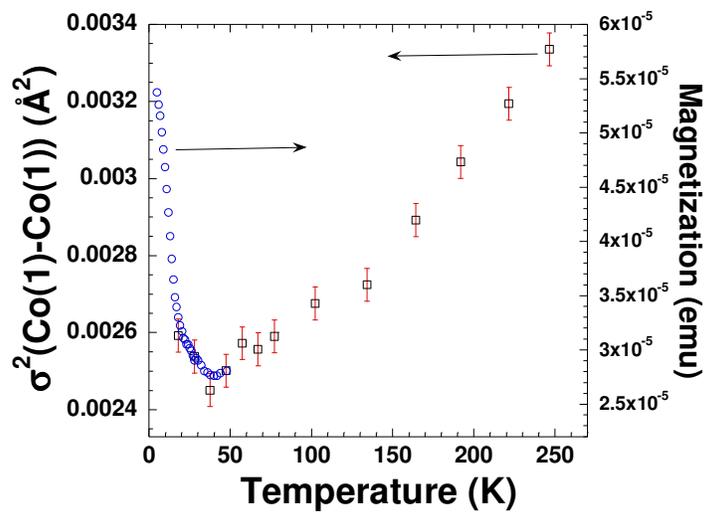

**Fig. 4.** Tyson *et al.*

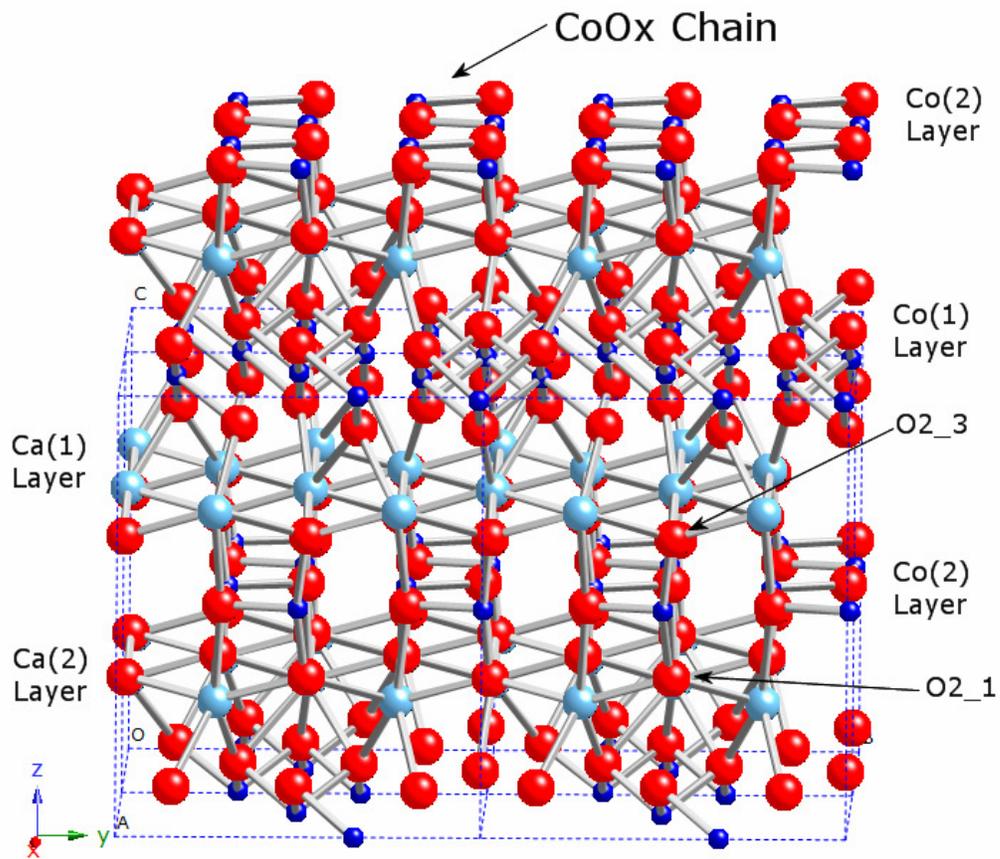